\documentclass[conference]{IEEEtran}
\IEEEoverridecommandlockouts
\usepackage{cite}
\usepackage{amsmath,amssymb,amsfonts}
\usepackage{algorithmic}
\usepackage{graphicx}
\usepackage{textcomp}
\usepackage{xcolor}

\usepackage{subfigure}
\usepackage{booktabs}
\usepackage{marvosym}
\usepackage{multirow}
\usepackage{caption}

\def\BibTeX{{\rm B\kern-.05em{\sc i\kern-.025em b}\kern-.08em
    T\kern-.1667em\lower.7ex\hbox{E}\kern-.125emX}}
\begin{document}

\title{A No-reference Quality Assessment Metric for Point Cloud Based on Captured Video Sequences}

\author{\IEEEauthorblockN{Yu Fan*\textsuperscript{1}, Zicheng Zhang*\textsuperscript{1}, Wei Sun\textsuperscript{1}, Xiongkuo Min\textsuperscript{1}, Ning Liu\textsuperscript{1}, \\ Quan Zhou\textsuperscript{2}, Jun he\textsuperscript{2}, Qiyuan Wang\textsuperscript{2}, and Guangtao Zhai\textsuperscript{1}}
\IEEEauthorblockA{\textit{Department of Electronic Engineering}, \textit{Shanghai Jiao Tong University}\textsuperscript{1}\\
\textit{Department of Video Cloud Technology},
\textit{Bilibili Inc} \textsuperscript{2}\\
Shanghai, China \\
\{fy-sky,zzc1998\}@sjtu.edu.cn}}

\maketitle

\begin{abstract}
Point cloud is one of the most widely used digital formats of 3D models, the visual quality of which is quite sensitive to distortions such as downsampling, noise, and compression. To tackle the challenge of point cloud quality assessment (PCQA) in scenarios where reference is not available, we propose a no-reference quality assessment metric for colored point cloud based on captured video sequences.  Specifically, three video sequences are obtained by rotating the camera around the point cloud through three specific orbits. The video sequences not only contain the static views but also include the multi-frame temporal information, which greatly helps understand the human perception of the point clouds. Then we modify the ResNet3D as the feature extraction model to learn the correlation between the capture videos and corresponding subjective quality scores. The experimental results show that our method outperforms most of the state-of-the-art full-reference and no-reference PCQA metrics, which validates the effectiveness of the proposed method. 
\end{abstract}


\section{Introduction}
Point clouds, which are widely utilized to represent 3D contents, have played a vital role in immersive applications such as virtual reality \cite{xiong2021augmented}, mesh representation\cite{9722570}, 3D reconstruction \cite{kang2020review}, and metaverse \cite{ning2021survey}. However, limited by the storage space and transmission bandwidth, point clouds inevitably undergo lossy processes such as compression and simplification. Such processes may sacrifice quality-aware information to compromise with the bit rates. Additionally, affected by the sensor accuracy and rendering techniques, unintended distortions like noise and blur might damage the visual quality of the point clouds as well. Therefore, mechanisms that can effectively quantify the distortion of point clouds are urgently needed to provide useful guidelines for compression systems and improve the Quality of Experience (QoE) for viewers\cite{min2021screen,zhang2021full,zhang2022nosuper}.

\let\thefootnote\relax\footnotetext{ $^*$The authors marked with * make equal contributions to this work.}

According to the involving extent of the reference information, objective quality assessment can be divided into full-reference (FR), reduced-reference (RR), and no-reference (NR) metrics \cite{zhai2020per,zhang2021nolow,sun2019mc360iqa}. The latest FR point cloud quality assessment (FR-PCQA) metrics usually employ both geometry and color features for evaluation. PointSSIM \cite{alexiou2020pointssim} compares difference of local topology and color distributions between the reference and distorted point clouds. GraphSIM \cite{yang2020graphsim} infers point cloud quality through graph similarity. To better utilize the features, PCQM \cite{meynet2020pcqm} introduces a weighted linear combination of curvature and color information for assessment. In this paper, we mainly focus on NR metrics because such metrics do not need reference information, thus having a wider range of applications\cite{7938348}. NR point cloud quality assessment (NR-PCQA) can then be categorized into projection-based methods and model-based methods. The former methods operate by analyzing the quality of 2D projections of the point clouds. For example, PQA-net \cite{liu2021pqa} identifies and evaluates the distortions by multi-view-based joint feature extraction. While the latter methods directly use the information in 3D domains and are invariant to projection parameters. Namely, 3D-NSS  \cite{zhang2022no,zhang2021nomesh} calculates both geometry and color attributes and employ well-behaving distribution models to quantify the distortions.  ResCNN \cite{liu2022point} designs an end-to-end sparse convolutional neural network to estimate the quality levels.

\begin{figure*}
    \centering
    \includegraphics[width=15cm]{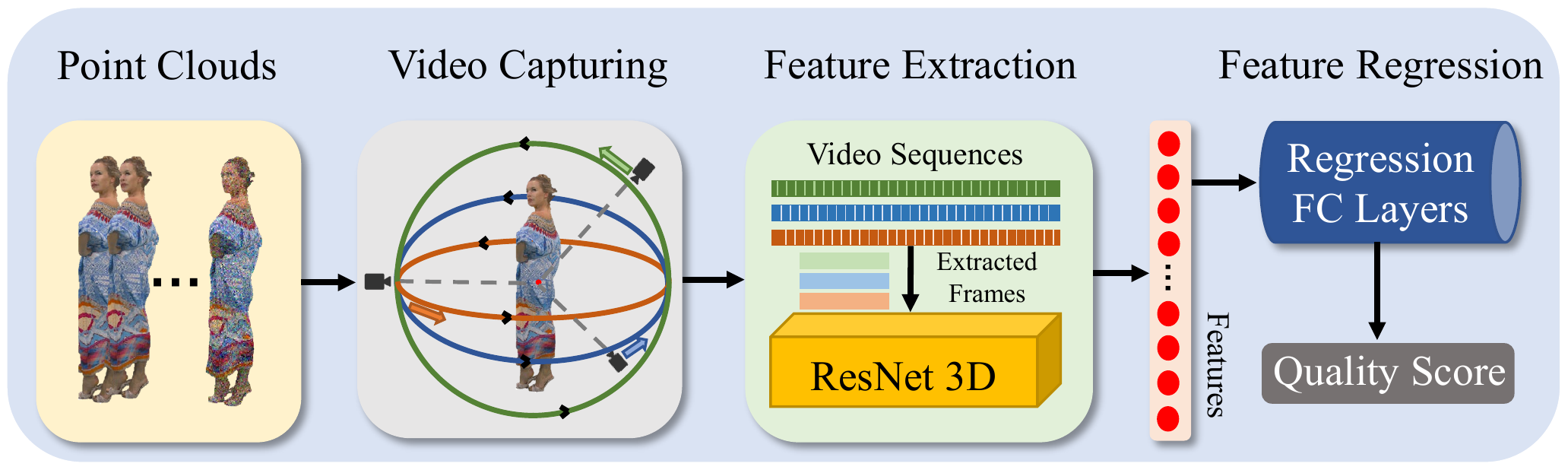}
    \caption{The framework of the proposed method.}
    \label{fig:framework}
    \vspace{-0.3cm}
\end{figure*}

\begin{figure}
    \centering
    \includegraphics[width = 6cm]{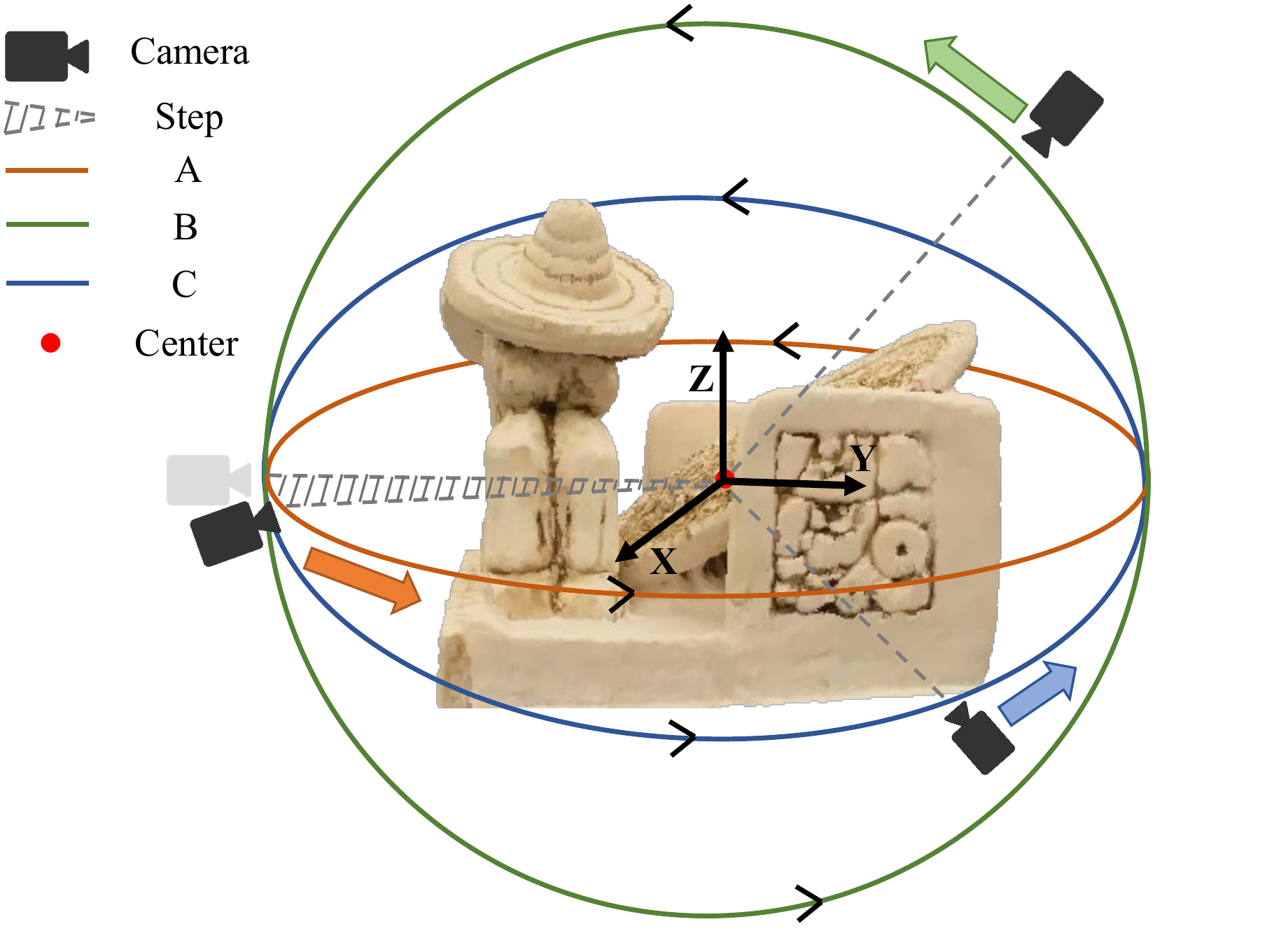}
    \caption{Illustration of the video sequences capturing process.}
    \label{fig:video}
    \vspace{-0.3cm}
\end{figure}

\begin{figure*}
    \centering
    \subfigure[]{\includegraphics[width = 17.5 cm]{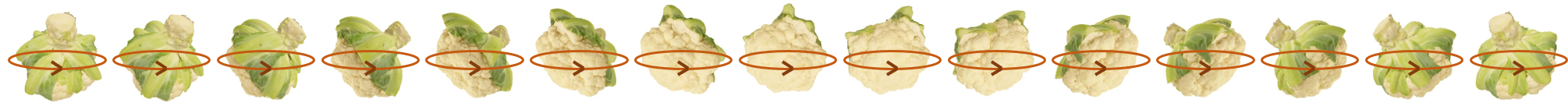}}
    \subfigure[]{\includegraphics[width = 17.5 cm]{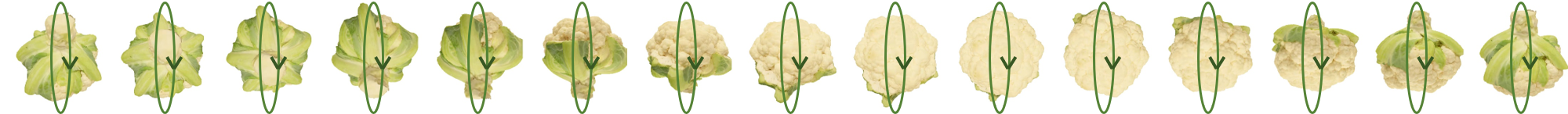}}
    \subfigure[]{\includegraphics[width = 17.5 cm]{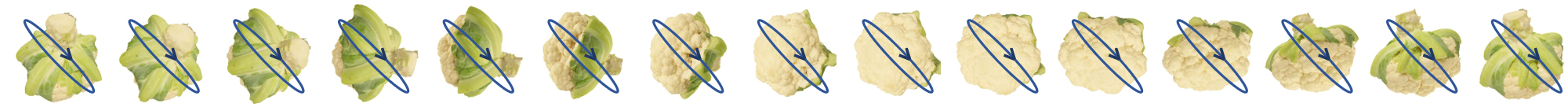}}
    \caption{Examples of captured videos sequences of the point cloud $Cautiflower$ from the WPC database, where (a), (b), and (c) represent the samples of frames captured from $OrbitA$, $OrbitB$, and $OrbitC$.}
    \label{fig:samples_frames}
    \vspace{-0.3cm}
\end{figure*}

Limited by the scale of previously proposed databases \cite{su2019perceptual,yang2020predicting}, model-based deep learning methods are hard to develop. The projection-based methods can increase the number of training samples by capturing more projections from different viewpoints, however, the projections are usually separate and static. Considering the success of video quality assessment \cite{sun2022deep}, we propose a novel no-reference quality assessment metric for point cloud based on captured video sequences. In some situations, people usually perceive point clouds through rotation rather than observing from static viewpoints. Captured video sequences not only contain static single-frame information but also include multi-frame temporal information, which can better help model understand human perception\cite{li2020subjective}. Specifically, the camera is rotated around the center of the point cloud through three designed orbits to capture three corresponding video sequences. Then sub-sequences of frames are randomly sampled from the original video sequences with certain intervals and a ResNet3D \cite{tran2018closer} model is applied to extract quality-aware features from the sub-sequences. Finally, the extracted features are regressed into quality levels through fully connected layers. In the experiment section, FR-PCQA, RR-PCQA, and NR-PCQA methods are employed for comparison. Additionally, to expand the range of competitors, several mainstream video quality assessment (VQA) methods are selected for validation as well. The experimental results and statistical comparison show that the proposed method achieves the best performance on both the Waterloo Point Cloud Database (WPC) \cite{su2019perceptual} and the SJTU-PCQA Database \cite{yang2020predicting} among the no-reference models, which indicates that the proposed framework can help improve the performance of PCQA methods and provide useful guidelines for point cloud processing algorithms.

\section{Proposed Method}
The framework of the proposed method is exhibited in Fig. \ref{fig:framework}, which includes the video capturing module, the feature extraction module, and the feature regression module.
\subsection{Videos Sequences Capture}
\label{sec:capture}
Given a point cloud $P$, we obtain the corresponding video sequences $PV$ with the Python package open3d \cite{open3d}:
\begin{equation}
    PV = Capture(P),
\end{equation}
where $Capture(\cdot)$ denotes the video generation process. Specifically, the camera is first placed at the default position calculated by the visualization function of open3d. Then the mean center of the point cloud $P$ can be derived as:
\begin{align}
    C_{\alpha} &  = \frac{1}{N} \sum_{i=1}^{N} \alpha_{i}, \\
    \alpha & \in \{X,Y,Z\},
\end{align}
where $N$ indicates the number of the points, $C_{\alpha}$ stands for the $X,Y,Z$ coordinates of the point cloud's mean center, and $\alpha_{i}$ denotes the $X,Y,Z$ coordinates of each point in the point cloud. Then we rotate the camera through three orbits to capture the video sequences and Fig. \ref{fig:video} illustrates the details of the capturing process. We move the origin of the coordinate system to the center of the point cloud and define the coordinates of camera position as ($X_{cam},Y_{cam},Z_{cam}$). To cover the viewpoints as much as possible, we define three circle orbits for video capturing, which can be derived as:
\begin{equation}
\boldsymbol{Orbit A:}\left\{
\begin{aligned}
X_{cam}^2 + Y_{cam}^2 &= R^2, \\
Z_{cam} &= 0, 
\end{aligned}
\right.
\end{equation}
\begin{equation}
\boldsymbol{Orbit B:}\left\{
\begin{aligned}
Y_{cam}^2 + Z_{cam}^2 &= R^2, \\
X_{cam} &= 0, 
\end{aligned}
\right.
\end{equation}
\begin{equation}
\quad \quad  \quad \boldsymbol{Orbit C:}\left\{
\begin{aligned}
X_{cam}^2 + Y_{cam}^2 + Z_{cam}^2 &= R^2, \\
X_{cam} + Z_{cam} &= 0, 
\end{aligned}
\right.
\end{equation}
where $\boldsymbol{Orbit A}$, $\boldsymbol{Orbit B}$, and $\boldsymbol{Orbit C}$ represent the three corresponding orbits, $R$ indicates the radius of the circle. 

In order to ensure the consistency of the video sequences, the camera rotation step is set as 1.71$^\circ$ between consecutive frames. More specifically, the camera rotates 1.71$^\circ$ around the center on the corresponding orbit to capture the next frame after capturing the current frame. Therefore, a total of 360/1.71=210 frames are obtained to cover each circle orbit. For a point cloud, three video sequences are generated, which contain 630=210$\times$3 frames in total. Then we refer to the three video sequences of the given point cloud $P$ as $PV_{A}$, $PV_{B}$, and $PV_{C}$ respectively. The examples are shown in Fig. \ref{fig:samples_frames}, from which we can observe the frame samples of the point cloud $Cauliflower$ in the WPC database from three orbits.

\subsection{Feature Extraction}
In this section, we describe the process of extracting frames from each video sequence . During the training stage, a frame is randomly selected between the first and the seventh frame as the start frame. Then we extract the following frames at a interval of 7 frames and obtain a total of 210/7 = 30 frames as the input sequence $PV_{in}$. With the number of epochs increasing, we can make use of most video information and do not exceed the limitation of GPU memory at the same time. 

To capture the quality-aware features, we employ the ResNet3D \cite{tran2018closer} with 4 residual layers as the feature extraction model. ResNet3D utilizes 3D convolutions to extract features from videos and is capable of using both temporal and spatial information, which has been widely used in many vision tasks. Then the quality-aware features can be obtained through:
\begin{equation}
    F = ResNet3D(PV_{in}),
\end{equation}
where $F$ indicates the extracted features (we delete the final linear layer of ResNet3D and modify the output into a vector of 128 features), $ResNet3D(\cdot)$ denotes for the ResNet3D feature extraction module.

\subsection{Feature Regression}
After the feature extraction module, a fully-connected (FC) layer consisting of 128 neurons is used as the regression model. Additionally, the three video sequences $PV_{A}$, $PV_{B}$, and $PV_{C}$ are labeled with the same quality score of the point cloud in the training stage. The average score of the three video sequences is recorded as the predicted score in the testing stage. Then the predicted quality scores $\boldsymbol{Q}_{p}$ can be computed as:
\begin{equation}
    \boldsymbol{Q}_{p} = {\rm FC}(F),
\end{equation}
where $FC$ represents the fully connected layer. The mean squared error (MSE) is employed as the loss function, which can be derived as:
\begin{equation}
    Loss = || \boldsymbol{Q}_{p}- \boldsymbol{Q}_{l}||^2_{2}
\end{equation}
where $\boldsymbol{Q}_{l}$ are mean opinion scores (MOS) for the distorted point cloud samples.

\begin{table*}[h]
\centering
\renewcommand\arraystretch{1.5}
\caption{Performance results on the WPC and SJTU-PCQA databases.}
\begin{tabular}{c|c|c|cccc|cccc}
\toprule
\multirow{2}{*}{Index}&\multirow{2}{*}{Type}&\multirow{2}{*}{Methods} & \multicolumn{4}{c|}{WPC} & \multicolumn{4}{c}{SJTU-PCQA} \\ 
\cline{4-11}
        &  & & SRCC      & PLCC      & KRCC     & RMSE     & SRCC      & PLCC      & KRCC       & RMSE \\ \hline
A&\multirow{2}{*}{FR} &
  PCQM     &0.7434& 0.7499 &0.5601 &15.1639  & 0.8673   & 0.8879    & \textbf{0.7017}    & 1.0804                 \\
B & & GraphSIM  & 0.5831    & 0.6163    & 0.4194   & 17.1939   & \textbf{0.8723}    & \textbf{0.8981}    &  0.6904  &  \textbf{1.0327}  \\
C & & PointSSIM     & 0.4542& 0.4667& 0.3278& 20.2733   & 0.7199  & 0.7438  & 0.5479 & 1.5424 \\ \hline
D & \multirow{1}{*}{RR} &
 PCMRR    & 0.3097 & 0.3433 & 0.2082 & 21.5302   & 0.4745  & 0.6836  & 0.3372 & 1.7246 \\ \hline
E & \multirow{4}{*}{NR} &
PQA-net      & 0.6900    & 0.7000    & 0.5100   & 15.1800   & -   &-    & - & -     \\
F & &3D-NSS      & 0.6479    & 0.6514    & 0.4417   & 16.5716   & 0.7144 & 0.7382  & 0.5174 & 1.7686   \\
G & &BRISQUE & 0.2746    & 0.3062  & 0.1933 & 21.8651 & 0.3923    & 0.4203  & 0.2971 & 2.0954  \\

H & &VSFA & 0.6274   & 0.6220    & 0.4501 & 17.0966 & 0.7168    &0.8140 & 0.5387     &  1.5018  \\
I & &StairVQA & 0.7234 &0.7179 & 0.5274 & 15.0749 & 0.7940 & 0.7858 & 0.5535 & 1.4264\\
J& & Proposed & \textbf{0.7558}    & \textbf{0.7679}    & \textbf{0.5643}  & \textbf{13.5605}   & 0.8320   & 0.8612   & 0.6045 & 1.2234  \\

                      \bottomrule
\end{tabular}
\label{tab:experiment}
\end{table*}

\begin{figure*}
  \centering
  \subfigure[]{\includegraphics[height=5cm]{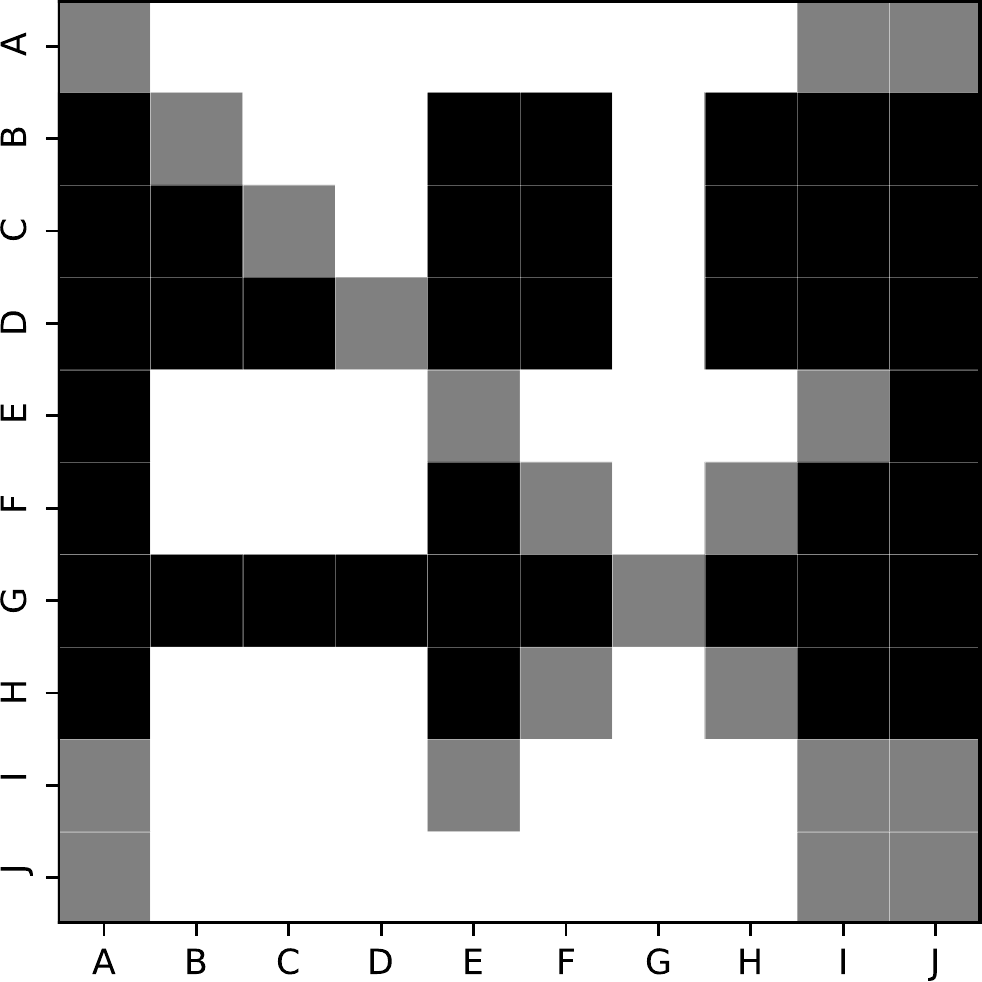}}
  \hspace{2cm}
  \subfigure[]{\includegraphics[height=5cm]{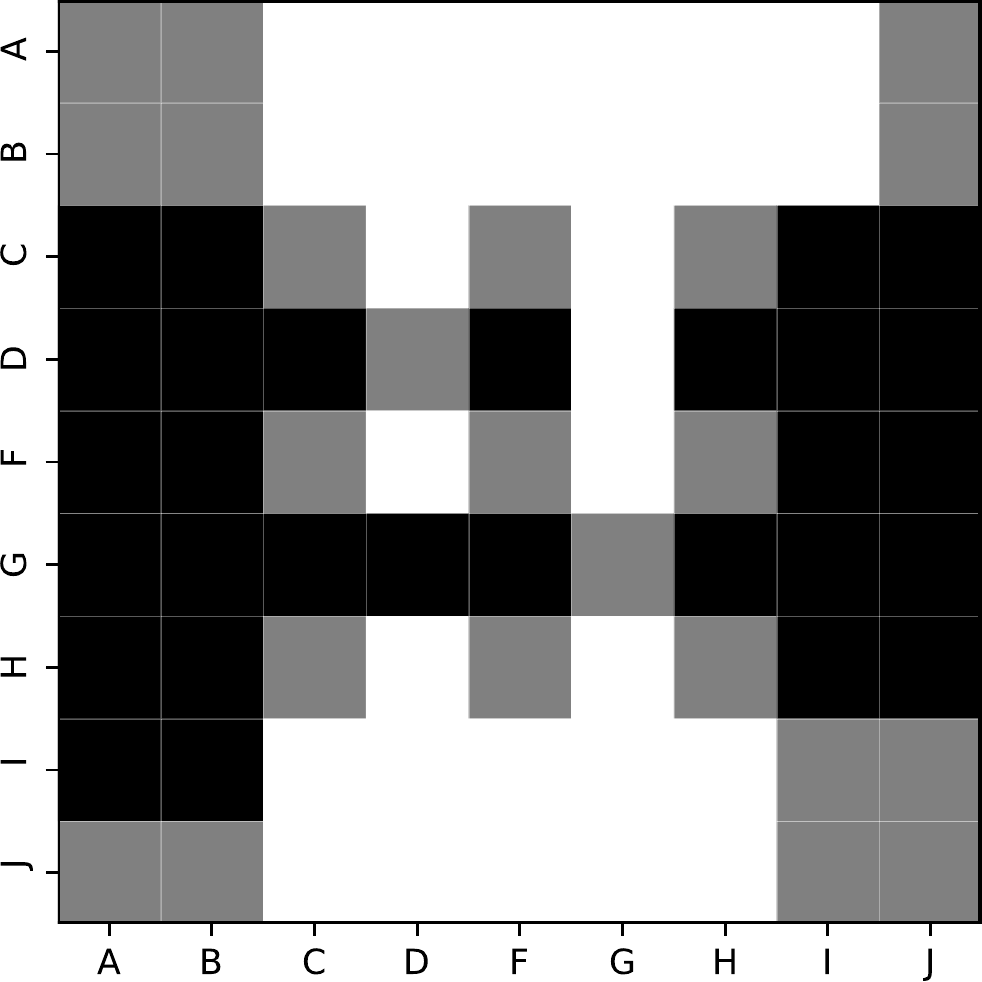}}
  \caption{Statistical significance test results on WPC database and JTU-PCQA database. A white/black block indicates that the row model is statistically
better/worse than the column model. A gray block indicates that the row and column models are statistically indistinguishable. A-J are model indices given
in Tables I. (a) WPC database. (b) SJTU-PCQA database.}
  \label{fig:statistical}
\end{figure*}


\section{Experiment}
\subsection{Validated Databases}
We select the Waterloo Point Cloud Database (WPC) \cite{su2019perceptual} and the SJTU-PCQA Database \cite{yang2020predicting} as the validation databases. The WPC database contains 20 reference point clouds and each reference point cloud is processed with five distortion types at different quality levels, which generates 740 distorted point clouds. The SJTU-PCQA database has 9 source point clouds and each point cloud is processed with seven distortion types at six quality levels, thus obtaining 378 distorted point clouds.


\subsection{Experiment Setup}
In this section, we discuss the details of the experiment. For the WPC database, we randomly split the training set and the testing set with a ratio of 8:2 10 times with no overlap. The average results are recorded as the final performance. For the SJTU-PCQA database, we do a 9-folds cross validation, select 8 source point clouds' distorted point clouds as the training set and leave the rest as the testing set. Such a process is repeated 9 times to ensure every group of point clouds has been used as the testing set. The average performance results are recorded as the final experimental results. During the training process, the Adam optimizer is employed with the initial learning rate set as 1e-4 and the batch size is set as 4. The input frames are first resized into 520x520 resolution and then center-cropped into 448x448 patches. Additionally, the number of training epochs is set as 50

\subsection{Experiment Criteria}
Four mainstream consistency evaluation criteria are utilized to compare the correlation between the predicted scores and MOSs, which include Spearman Rank Correlation Coefficient (SRCC), Kendall’s Rank Correlation Coefficient (KRCC), Pearson Linear Correlation Coefficient (PLCC), Root Mean Squared Error (RMSE). 
Specifically, SRCC is used to measure the correlation of ranks, PLCC denotes the linear correlation, KRCC describes the similarity of the orderings, and RMSE indicates the accuracy of the predicted scores. It's worth mentioning that an excellent model should obtain values of SRCC, KRCC, PLCC close to 1 and RMSE to 0 .


\subsection{Experiment Competitors}
To demonstrate the effectiveness of the proposed method, several FR-PCQA and NR-PCQA methods are chosen for comparison. Additionally, we also select some well-known NR-VQA metrics for further validation. The methods are as follows:
\\ $\bullet$  FR-PCQA models:  FR metrics consist of PCQM \cite{meynet2020pcqm}, GraphSIM \cite{yang2020graphsim}, and PointSSIM \cite{alexiou2020pointssim}. Note that the FR-PCQA methods are validated on the same testing set to make the comparison even.
\\ $\bullet$  RR-PCQA models: We select PCMRR \cite{viola2020pcmrr} as the RR-PCQA metric for comparison. 
\\ $\bullet$  NR-PCQA models: These metrics include PQA-net \cite{liu2021pqa}, 3D-NSS \cite{zhang2021no}, BRISQUE \cite{mittal2012brisque}, VSFA \cite{li2019vsfa}, and StairVQA \cite{sun2021deep}.  It's worth mentioning that PQA-net is developed for point cloud for single type of distortion, therefore we do not validate PQA-net on the SJTU-PCQA database, which includes mixed distortions. Besides, the VQA-based metrics (BRISQUE, VSFA, and StairVQA) share the same experiment setup as the proposed method.

\subsection{Experiment Performance}
The final experiment results are exhibited in Table \ref{tab:experiment}. The best performance for each database is marked in bold.  The proposed method achieves first place on the WPC database and obtains a slight gap with the FR-PCQA metrics on the SJTU-PCQA database.

To testify whether the results between different models are statistically significant, we adopt a t-test on the SRCC values\cite{sheskin2003handbook} produced by the chosen models. The results of comparison between every pairs are summarized in Fig. 4, where a white/black block indicates the row model is statistically better/worse than the column model, and a gray block indicates the row model has no statistical difference with the column model. It can be found that our proposed method achieves statistically similar performance to the current best FR PCQA methods on both WPC database and SJTU-PCQA database.

With closer inspections, several observations can be made. 1) All the PCQA models have a clear performance drop on the WPC database. We attempt to give the reasons. The WPC database is more diverse in content and includes more distorted point clouds, which may challenge the effectiveness of the PCQA models. 2) With the information of reference point clouds, the leading FR-PCQA metrics are generally superior to NR-PCQA metrics. Surprisingly, the proposed method achieves indistinguishable performance with the most effective FR-PCQA method PCQM and outperforms all the compared NR-PCQA methods. The reasons are as follows. We employ video sequences for analysis, which cover more quality-aware viewpoints than most projection-based methods. For example, the PQA-net uses 12 static viewpoints while the proposed method utilizes 3 video sequences containing 630 frames, which help improve the performance. The proposed framework makes better use of the temporal information with the assistance of ResNet3D. We think that in some situations, people perceive the point clouds mostly by rotating rather than simply observing from static viewpoints, which makes our method gain more advantage.

\section{conclusion}
To deal with the PCQA tasks, this paper proposes a novel framework that operates on the captured video sequences. The videos are captured by rotating the camera around the point cloud through 3 fixed orbits and features are extracted from the videos using ResNet3D. In this way, the proposed method can make use of not only static single-frame information, but also multi-frame temporal information, which significantly improves the performance of the projection-based methods. The experimental results and statistical comparison show that our method outperforms most of the state-of-the-art full-reference and no-reference PCQA metrics on the  WPC and SJTU-PCQA databases, which validates the effectiveness of the proposed method. 

\bibliographystyle{IEEEbib}
\bibliography{mybib}


\end{document}